\documentclass{jetpl}

\usepackage{amssymb}
\usepackage{amsmath}
\usepackage{graphicx}

\twocolumn \lat

\DeclareMathOperator{\sgn}{sgn}

\title{Superconducting triplet spin valve}
\rtitle{Superconducting triplet spin valve}
\sodtitle{Superconducting triplet spin valve}

\author{Ya.\,V.~Fominov$^{\,+}$,
A.\,A.~Golubov$^{\,*}$,
T.\,Yu.~Karminskaya$^{\,\nabla}$,
M.\,Yu.~Kupriyanov$^{\,\nabla}$,
R.\,G.~Deminov$^{\,\square}$,
L.\,R.~Tagirov$^{\,\square}$}

\rauthor{Ya.\,V.~Fominov, A.\,A.~Golubov, T.\,Yu.~Karminskaya, et al.}

\sodauthor{Fominov, Golubov, Karminskaya, et al.}

\address{$^+$ L.\,D.~Landau Institute for Theoretical Physics RAS, 119334 Moscow, Russia\\
$^*$ Faculty of Science and Technology and MESA+ Institute of Nanotechnology,
University of Twente, P.O.\ Box 217, 7500 AE Enschede, The Netherlands\\
$^\nabla$ Nuclear Physics Institute, Moscow State University, 119992 Moscow, Russia\\
$^\square$ Physics Faculty, Kazan State University, 420008 Kazan, Russia}

\dates{25 May 2010}{*}

\abstract{We study the critical temperature $T_c$ of SFF trilayers (S is a singlet superconductor,
F is a ferromagnetic metal), where the long-range triplet superconducting component is generated
at noncollinear magnetizations of the F layers. We demonstrate that $T_c$ can be a nonmonotonic
function of the angle $\alpha$ between the magnetizations of the two F layers. The minimum is achieved
at an intermediate $\alpha$, lying between the parallel (P, $\alpha=0$) and antiparallel (AP, $\alpha=\pi$)
cases. This implies a possibility of a ``triplet'' spin-valve effect: at temperatures
above the minimum $T_c^\mathrm{Tr}$ but below $T_c^\mathrm{P}$ and $T_c^\mathrm{AP}$, the system is superconducting only in the vicinity of the collinear orientations. At certain parameters, we predict a reentrant $T_c(\alpha)$ behavior.
At the same time, considering only the P and AP orientations, we find that both the
``standard'' ($T_c^\mathrm{P} <T_c^\mathrm{AP}$) and ``inverse'' ($T_c^\mathrm{P} >T_c^\mathrm{AP}$) switching effects are possible depending
on parameters of the system.\\ \centerline{\textbf{Published as: JETP Letters \textbf{91}, 308 (2010) [Pis'ma v
ZhETF \textbf{91}, 329 (2010)]}}}


\begin{document}

\maketitle

In superconducting spin valves with the layer sequence F1/S/F2 the superconducting transition
temperature $T_c$ of the system can be controlled by mutual alignment of magnetizations $\mathbf{M}_{1,2}$
of the two ferromagnetic layers F1 and F2. Therefore, at a temperature $T$ fixed inside the range of $T_c$
variation, there is an opportunity for switching the superconductivity on and off by reversing the
magnetization direction of the F1 or F2 layer. Model calculations have shown that the transition
temperature $T_c^\mathrm{AP}$ for the antiparallel ($\mathbf{M}_{1}\uparrow \downarrow \mathbf{M}_{2}$)
orientation of the F1 and F2 magnetizations should be higher than the transition temperature $T_c^\mathrm{P}$
for the opposite case ($\mathbf{M}_{1}\uparrow \uparrow \mathbf{M}_{2}$) \cite{Tagirov,Buzdin1,Fominov1}.
The situation with this order of $T_c$'s (i.e., $T_c^\mathrm{P} < T_c^\mathrm{AP}$) is commonly referred to as the ``standard''
switching (see, e.g., \cite{Moraru2}), and the switching in this case actually occurs at temperatures $T$
such that $T_c^\mathrm{P} < T< T_c^\mathrm{AP}$. The basic physical reason for the difference
$\Delta T_c =T_c^\mathrm{AP}-T_c^\mathrm{P}>0$ is partial compensation of the pair-breaking ferromagnetic
exchange field, if the magnetizations of the F1 and F2 layers are aligned antiparallel.

Several experimental groups have published results on superconducting spin valves of the F1/S/F2
type \cite{Moraru2,Gu,Potenza,Moraru1,Nowak,Rusanov,Steiner,Singh1,Kim,Leksin}. The experimental results
turned out to be controversial. Some studies of F1/S/F2 structures have shown the standard spin-valve
effect \cite{Moraru2,Gu,Potenza,Moraru1,Nowak} with the maximum shift $\Delta T_c\approx 41$\,mK reported
for the Ni/Nb/Ni trilayer in \cite{Moraru1}. However, some experiments revealed the ``inverse''
spin-valve effect \cite{Rusanov,Steiner,Singh1,Leksin} with $T_c^\mathrm{P}>T_c^\mathrm{AP}$ (i.e., $\Delta T_c <0$).
The most advanced calculations within the proximity effect theory, which take into account the
triplet components of the superconducting pairing \cite{Bergeret}, demonstrate only the standard
switching \cite{Fominov1,Linder} with $T_c$ monotonically increasing from the P to AP
configuration \cite{Fominov1}. Additional physical mechanisms like spin imbalance
effect \cite{Rusanov,Singh1} or magnetic domain structure \cite{Steiner,Kim} should be recruited
to explain the inverse spin-valve effect in the studied F1/S/F2-type structures.

A bit earlier an unconventional spin-valve-like S/F1/F2 structure was theoretically proposed
in \cite{Oh} to control the superconducting $T_c$ in the S layer by mutual alignment of the magnetizations
of the two \emph{adjacent} ferromagnetic layers F1 and F2. The authors of \cite{Oh} argued that $T_c^\mathrm{P}<T_c^\mathrm{AP}$ in their
system because of partial cancelation of the pair-breaking exchange fields just within
the magnetic F1/F2 subsystem of the structure, thus predicting the standard switching as
in the interleaved F1/S/F2 structure.

The S/F1/F2 structures are much less investigated experimentally \cite{Nowak,Westerholt},
and the experiments indicate the standard switching effect \cite{Oh} with the maximal size of about 200\,mK.
\begin{figure}[t]
 \centerline{\includegraphics[width=82mm]{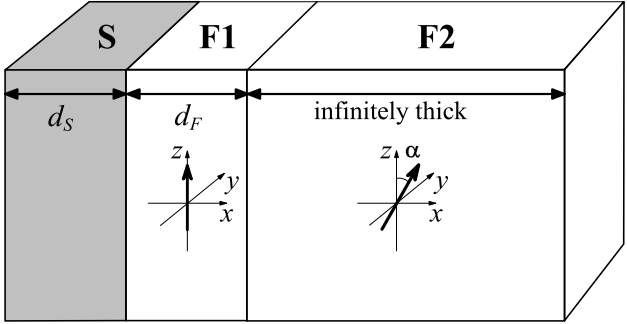}}
\caption{Fig.~\ref{fig:fig1}. S/F1/F2 trilayer. The S/F1 interface corresponds to $x=0$.
The thick arrows in the F layers denote the exchange fields $\mathbf{h}$ lying in the $(y,z)$ plane.
The angle between the in-plane exchange fields is $\alpha$.}
 \label{fig:fig1}
\end{figure}
In this Letter we study the critical temperature of a S/F1/F2 trilayer at arbitrary angle
between the in-plane magnetizations of the ferromagnetic layers (see Fig.~l).
We demonstrate that this structure allows not only the standard but also inverse
spin-switching effect. Moreover, we show for the first time that the minimal critical temperature $T_c^\mathrm{Tr}$ of the structure is achieved at a noncollinear alignment of the magnetizations, when the long-range triplet component of the superconducting pairing
is generated. Since $T_c^\mathrm{Tr}$ is lower than both $T_c^\mathrm{P}$ and $T_c^\mathrm{AP}$, this offers a possibility of a ``triplet spin-valve effect'' never reported before.

\textbf{1.~Model.} We consider the S/F1/F2 structure in the dirty limit, which is described
by the Usadel equations. Near $T_c$, the Usadel equations are linearized and contain only
the anomalous Green function $\check{f}$ \cite{Bergeret,Ivanov}:
\begin{equation} \label{eq1}
\frac{D}{2} \frac{d^{2}\check{f}}{dx^{2}}-|\omega |\check{f}-\frac{i\sgn\omega}{2}
\left\{ \hat{\tau}_{0}(\mathbf{h}\hat{\boldsymbol\sigma}),\check{f}
\right\} +\Delta \hat{\tau}_{1}\hat{\sigma}_0 =0.
\end{equation}
Here, $\check{f}$ is a 4$\times$4 matrix, $\hat{\tau}_{i}$ and $\hat{\sigma}_{i}$ are the Pauli
matrices in the Nambu-Gor'kov and spin spaces, respectively, $D$ is the diffusion constant,
and $\omega = \pi T_c (2n+1)$ with integer $n$ is the Matsubara frequency. The exchange field
in the middle F1 layer is along the $z$ direction, $\mathbf{h} = (0,\ 0,\ h)$, while the
exchange field in the outer F2 layer is in the $yz$ plane:
$\mathbf{h} =\left( 0,\ h\sin \alpha ,\ h\cos \alpha \right)$.
The angle $\alpha$ changes between 0 (parallel configuration, P) and $\pi$ (antiparallel configuration, AP).
The order parameter $\Delta$ is real-valued in the superconducting layer, while in the
ferromagnetic layers it is zero.
In general, the diffusion constant $D$ acquires a proper subscript, S or F, when Eq.\ (\ref{eq1})
is applied to the superconducting or ferromagnetic layers, respectively. However, for simplicity we take them equal in this paper, because this assumption does not influence qualitative behavior of $T_c(\alpha)$.

The Green function $\check{f}$ can be expanded into the following components:
\begin{equation} \label{eq2}
\check{f}=\hat{\tau}_{1}\left( f_{0}\hat{\sigma}_{0}+f_{3}\hat{\sigma}_{3}+f_{2}\hat{\sigma}_{2}\right) ,
\end{equation}
where $f_{0}$ is the singlet component, $f_{3}$ is the triplet with zero projection on the $z$ axis,
and $f_{2}$ is the triplet with $\pm 1$ projections on $z$ (the latter is present only
if $\alpha \neq 0,\pi$). The singlet component is even in frequency (and real-valued), while the triplet ones are odd (and imaginary): $f_{0}(-\omega ) = f_{0}(\omega)$, $f_{3}(-\omega ) = -f_{3}(\omega)$, and $f_{2}(-\omega ) = -f_{2}(\omega)$,
which makes it sufficient to consider only positive Matsubara frequencies, $\omega >0$.

As we show below, the problem of calculating $T_c$ can be reduced to an effective set of equations for the
singlet component in the S layer: the set includes the self-consistency equation and the Usadel equation,
\begin{gather}
\Delta \ln \frac{T_{cS}}{T_{c}}=2\pi T_{c}\sum\limits_{\omega >0}\left(
\frac{\Delta }{\omega }-f_{0}\right) , \label{eq4} \\
\frac{D}{2}\frac{d^{2}f_{0}}{dx^{2}} -\omega
f_{0}+\Delta =0, \label{eq5}
\end{gather}
with the boundary conditions
\begin{equation} \label{eq6}
\left. \frac{df_{0}}{dx}=0\right|_{x=-d_{S}},\qquad \left. -\xi \frac{df_{0}}{dx}=Wf_{0} \right|_{x=0}.
\end{equation}
Here $T_{cS}$ and $\xi=\sqrt{D/2\pi T_{cS}}$ are the superconducting transition
temperature and coherence length for an isolated S layer, and we assume that the S layer occupies the region $-d_{S}<x<0$ (see Fig.~1). This is exactly the problem for which the multi-mode solution procedure (as well as the fundamental-solution method) was developed in \cite{Fominov2} and then applied to F1/S/F2 spin valves in \cite{Fominov1}. We only need to determine the explicit expression for $W$ in Eq.\ (\ref{eq6}), solving the boundary problem for the S/F1/F2 structure.

\textbf{2.~Solution of the model.} To simplify derivations, while keeping the essential
physics, we consider the middle ferromagnetic layer F1 of arbitrary
thickness ($0<x< d_{F}$) but the outer ferromagnetic layer F2 being
semi-infinite ($d_{F}<x<\infty $). The Usadel equation (\ref{eq1}) generates
the following characteristic wave vectors:
\begin{equation} \label{eq7}
k_\omega =\sqrt{\frac{2\omega}D},\quad k_h =\sqrt{\frac hD},\quad \tilde k_h =\sqrt{k_\omega^2 +2ik_h^2}.
\end{equation}
Only $k_\omega$ appears in the solution for the S layer, while the F-layers' solutions are described by $k_\omega$, $\tilde k_h$, and $\tilde k_h^*$. Since the exchange energy is usually larger than the superconducting energy scale, $h\gg T_c$,
the $k_\omega$ mode in the ferromagnetic layers (arising at noncollinear magnetizations) represents the \emph{long-range} triplet component \cite{Bergeret}, which plays the key role in the present study.

In the S layer the solution of Eq.\ (\ref{eq1}) is:
\begin{equation} \label{eq8}
\begin{pmatrix}
f_{0}(x) \\
f_{3}(x) \\
f_{2}(x)
\end{pmatrix}
=
\begin{pmatrix}
f_{0}(x) \\
0 \\
0
\end{pmatrix}
+
\begin{pmatrix}
0 \\
A \\
B
\end{pmatrix}
\frac{\cosh \left( k_\omega ( x+d_S) \right)}{\cosh \left( k_\omega d_{S}\right)}.
\end{equation}
The singlet component $f_0(x)$ in the S layer cannot be written explicitly,
since it is self-consistently related to the (unknown) order parameter $\Delta(x)$
by Eqs.\ (\ref{eq4})-(\ref{eq5}). Our strategy now is to obtain the effective boundary
conditions (\ref{eq6}) for $f_0(x)$, eliminating all other components in the three layers.

In the middle F1 layer the solution of Eq.\ (\ref{eq1}) reads:
\begin{gather}
\begin{pmatrix}
f_{0}(x) \\
f_{3}(x) \\
f_{2}(x)
\end{pmatrix}
=C_{1}
\begin{pmatrix}
0 \\
0 \\
1
\end{pmatrix}
\cosh \left( k_\omega x\right) + S_{1}
\begin{pmatrix}
0 \\
0 \\
1
\end{pmatrix}
\sinh \left( k_\omega x\right) + \notag \\
+C_{2}
\begin{pmatrix}
1 \\
1 \\
0
\end{pmatrix}
\cosh \left( \tilde{k}_{h}x\right) +C_{3}
\begin{pmatrix}
-1 \\
1 \\
0
\end{pmatrix}
\cosh \left( \tilde{k}_{h}^* x\right) + \notag \\
+S_{2}
\begin{pmatrix}
1 \\
1 \\
0
\end{pmatrix}
\sinh \left( \tilde{k}_{h}x\right)
+S_{3}
\begin{pmatrix}
-1 \\
1 \\
0
\end{pmatrix}
\sinh \left( \tilde{k}_{h}^* x\right) . \label{eq9}
\end{gather}

Finally, the solution in the semi-infinite, outer F2 layer is built only from descending modes:
\begin{align}
\begin{pmatrix}
f_{0}(x) \\
f_{3}(x) \\
f_{2}(x)
\end{pmatrix}
&= E_{1}
\begin{pmatrix}
0 \\
-\sin \alpha \\
\cos \alpha
\end{pmatrix}
\exp \left( -k_\omega ( x-d_F) \right) + \notag \\
&+ E_{2}
\begin{pmatrix}
1 \\
\cos \alpha \\
\sin \alpha
\end{pmatrix}
\exp \left( -\tilde{k}_{h} ( x-d_F) \right) + \notag \\
&+ E_{3}
\begin{pmatrix}
-1 \\
\cos \alpha \\
\sin \alpha
\end{pmatrix}
\exp \left( -\tilde{k}_{h}^* ( x-d_F) \right) . \label{eq10}
\end{align}

We will use the simplest, perfect-transparency boundary conditions at the S/F1 and F1/F2 interfaces
(the case $\gamma =1$ and $\gamma _{B}=0$ in the notations of~\cite{KL}):
\begin{equation} \label{eq11}
\left. f_i \right|_{\text{left}}=\left. f_i \right|_{\text{right}},\quad \left. \frac{df_i}{dx}\right|_{\text{left}}=\left.
\frac{df_i}{dx}\right|_{\text{right}}.
\end{equation}
Altogether there are 12 boundary conditions at the two interfaces (S/F1 and F1/F2). We are mainly interested in
one of them, determining the derivative of the singlet component on the S side of the S/F1 interface ($x=0$):
\begin{equation} \label{eq12}
\left. \frac{df_{0}}{dx}\right|_{x=0}=2\Real(\tilde{k}_{h} S_{2}).
\end{equation}
The remaining 11 boundary conditions form a system of 11 linear equations for 11 coefficients entering
Eqs.\ (\ref{eq8})-(\ref{eq10}). The solution of this system is nonzero due to $f_0(0)$
coming from Eq.\ (\ref{eq8}) and entering the ``right-hand side'' of the system. Finding
the $S_{2}$ coefficient [which is proportional to $f_{0}(0)$], we substitute it into Eq.\ (\ref{eq12})
and thus explicitly find $W$ entering the effective boundary conditions (\ref{eq6}).
\begin{figure}[t]
 \centerline{\includegraphics[width=75mm]{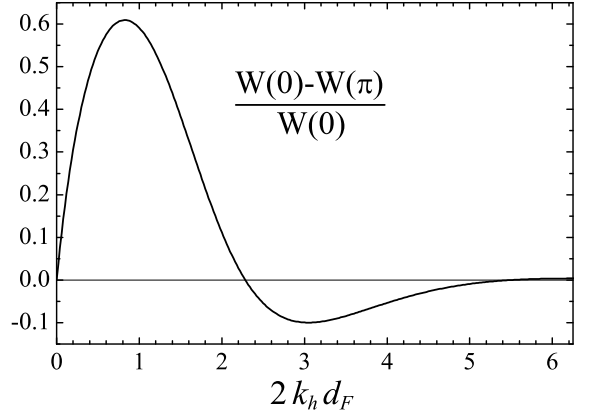}}
\caption{Fig.~\ref{fig:fig2}. Dependence of $W(0)-W(\pi)$, Eq.\ (\ref{eq13}), on the thickness $d_{F}$ of the F1 layer. Positive values of this oscillating function correspond to stronger suppression of superconductivity at the P alignment (the standard switching effect), while negative values correspond to stronger suppression of superconductivity at the AP alignment (the inverse switching effect).}
 \label{fig:fig2}
\end{figure}

\textbf{3. Analysis of the solution.} After reducing the problem to Eqs.\ (\ref{eq4})-(\ref{eq6}),
all the information about the two F layers is contained in the single real-valued function $W$.
This function makes $f_0(x)$ bend at the S/F1 interface, hence the larger $W$, the stronger $T_c$ is suppressed.

The explicit expression for $W(\alpha)$ is very cumbersome and we do not write it here.
However, certain analytical development (as well as complete numerical analysis) is possible.
For the analytical consideration, we make an additional assumption of $T_{c}\ll h$,
which implies $k_\omega \ll k_{h}$. For the collinear cases ($\alpha =0$ and $\alpha=\pi$)
we then find $W(0)=2k_h \xi$ and
\begin{equation} \label{eq13}
W(0)-W(\pi)=2k_{h} \xi \frac{\sqrt{2}\sin (2k_{h}d_{F}+\pi /4)-e^{-2k_{h}d_{F}}}{\sinh (2k_{h}d_{F})+\cos (2k_{h}d_{F})},
\end{equation}
which oscillates as a function of $d_{F}$, changing its sign (see Fig.~2).
Thus, as a result of interference in the middle F1 layer, we can either have the standard
spin-switching effect with $T_{c}^\mathrm{P}<T_{c}^\mathrm{AP}$ [when the pair-breaking
at the P alignment is stronger than at the AP alignment of magnetizations,
i.e., at $W(0)-W(\pi )>0$ as in the range $2k_{h}d_{F}<3\pi /4$ in Fig.~2] or the inverse
spin-switching effect with $T_{c}^\mathrm{P} >T_{c}^\mathrm{AP}$ [at $W(0)-W(\pi )<0$ as in
the range $3\pi /4<2k_{h}d_{F}<7\pi /4$ in Fig.~2]. Note that the amplitude of the inverse
effect is notably smaller compared with the standard one. The analytical calculation of
the second derivatives of $W(\alpha)$ at $\alpha =0$ and $\pi$ (the first ones are zero)
shows that under the above assumption, both the collinear alignments represent local minima
of $W(\alpha)$. This means that $T_{c}(\alpha )$ \emph{decreases} as the configuration
deviates from the P or AP alignment. Therefore, $T_{c}(\alpha )$ is nonmonotonic,
and the minimal $T_{c}$ must be achieved at some noncollinear configuration of
magnetizations at $\alpha \neq 0,\pi$.

\begin{figure}[t]
 \centerline{\includegraphics[width=82mm]{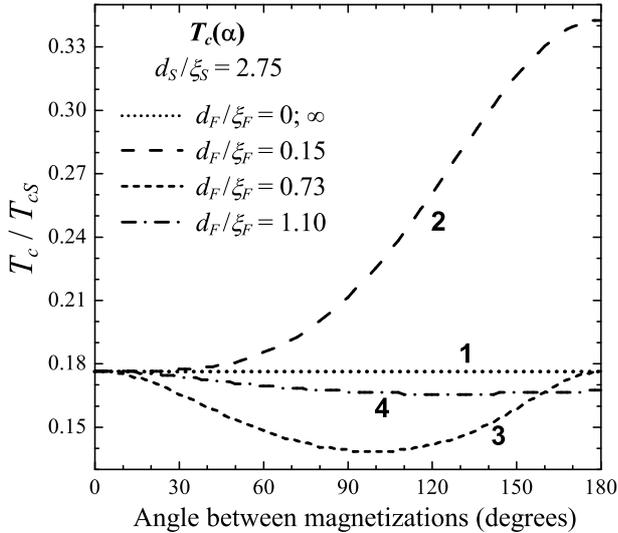}}
\caption{Fig.~\ref{fig:fig3}. Critical temperature $T_{c}$ vs.\ the misalignment angle $\alpha$ for various thicknesses of the F1 layer. We took $h/\pi T_{cS}=6.8$; all other parameters are shown in the figure. In the cases $d_F=0$ and $d_F=\infty$, which are physically equivalent (curve ``1''), $T_c$ does not depend on $\alpha$. Curves ``2'' and ``4'' correspond to the standard and inverse switching effects, respectively. Curve ``3'' demonstrates the triplet spin-valve effect. The coherence lengths $\xi_S$ and $\xi_F$ were taken equal (denoted by $\xi$ in the text) in order to present our main results in the simplest possible case. At $\alpha=0$ all the curves coincide, since in this case the F part of the system is physically equivalent to a single half-infinite F layer.}
 \label{fig:fig3}
\end{figure}

\begin{figure}[t]
 \centerline{\includegraphics[width=82mm]{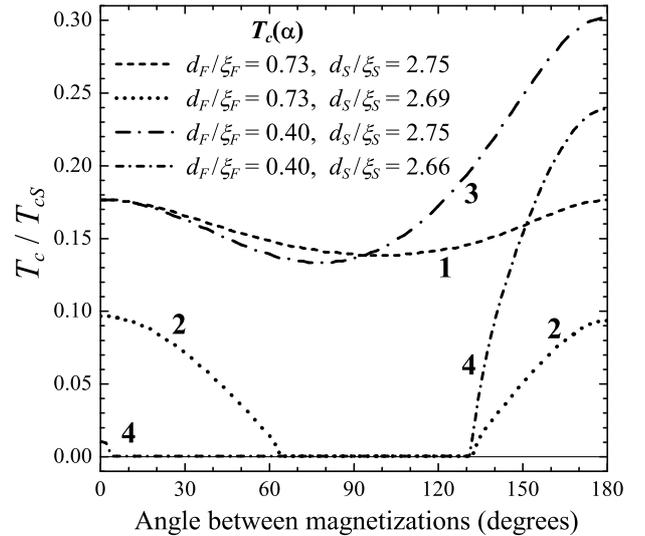}}
\caption{Fig.~\ref{fig:fig4}. $T_{c}(\alpha)$ at various $d_F$ and $d_S$. Curve ``1'' coincides with curve ``3'' in Fig.~3. Curves ``2'' and ``4'' demonstrate the reentrant behavior, in which case the triplet spin-valve effect takes place even at $T=0$.}
 \label{fig:fig4}
\end{figure}

The analytical results obtained at $k_\omega \ll k_h$ are illustrated and extended by
numerical calculations at arbitrary relation between $k_\omega$ and $k_{h}$. Figure~3
shows dependence of the transition temperature $T_{c}$ on the angle $\alpha$ between the magnetizations.
We see that at small thicknesses $d_{F}$ of the middle ferromagnetic layer F1, the switching
effect is standard, while at larger $d_{F}$ the effect is inverse ($T_{c}^\mathrm{P}>T_{c}^\mathrm{AP}$).
Moreover, when the F1 layer thickness is around a half of the coherence length $\xi$,
the minimal critical temperature $T_{c}^\mathrm{Tr}$ at noncollinear orientations is significantly
lower that both $T_{c}^\mathrm{P}$ and $T_{c}^\mathrm{AP}$~--- this case corresponds to the triplet spin-valve
effect. Note that depending on the parameters of the system, the minimum of $T_c(\alpha)$,
predicted analytically, can shift to a close vicinity of either $\alpha=0$ or $\alpha=\pi$,
becoming shallow and indistinguishable.

Figure 4 demonstrates the possibility of reentrant $T_c(\alpha)$ dependence. In this situation the triplet spin-valve effect takes place even at $T=0$.

\textbf{4.~Discussion.} The physical interpretation of the triplet spin-valve effect
can be given as follows: at the collinear configurations, both the singlet component
$f_{0}$ and the zero-projection triplet component $f_{3}$ of the pairing function are
short-ranged (with the characteristic penetration depth of the order of $k_{h}^{-1}$),
so that at $k_{h}^{-1}\ll d_{F}$ the middle F1 layer plays a role of a shield separating the S layer from
the ferromagnetic half-space F2. When the angle between magnetizations declines
from the collinear configurations, the long-range triplet component $f_{2}$ of the
pairing function is generated \cite{Bergeret}. Then, the S layer becomes effectively
coupled by this long-range triplet component to the semi-infinite ferromagnetic F2 layer.
The pair-breaking in the S layer enhances, giving rise to more effective suppression
of superconducting $T_{c}$. In other words, we can say that the $T_c$ suppression is
due to ``leakage'' of Cooper pairs into the ferromagnetic part. In this language,
the generation of the long-range triplet component opens up an additional channel
for this ``leakage'', hence $T_c$ is suppressed stronger.

In order to supply the qualitative picture by more quantitative details,
we can find the amplitudes of different components at the S/F1 interface
in the limit of $k_\omega \ll k_h$ and large $d_F$. This can be done analytically
from the boundary conditions which produce the linear system of equations for the
coefficients entering Eqs.\ (\ref{eq8})-(\ref{eq10}). We find that in the limit of $d_F\gg k_\omega^{-1},k_h^{-1}$, the amplitudes of the long-range triplet components near the S/F1
interface (which are given by $C_1$, $S_1$, and $B$) are suppressed by the
factor $e^{-k_\omega d_F - k_h d_F}$ which has a clear physical interpretation.
The long-range components are generated from the short-range ones at the F1/F2
interface (i.e., at $x=d_F$), where electrons ``feel'' inhomogeneous magnetization.
Therefore, the long-range contribution at the S/F1 interface is obtained as a
result of a ``wave'' that goes from the S/F1 interface as a short-range component
with the wave vector $k_h$ and returns after reflection at the F1/F2
interface as a long-range component with the wave vector $k_\omega$.
At the same time, the self-consistency equation (\ref{eq4}) that determines
$T_c$, contains only the singlet short-range component. Therefore, the influence
of the long-range components on $T_c$ is indirect: the long-range components
influence $T_c$ only through their influence on the singlet component.
We find that while the difference between $W$ (that encodes the information about the suppression
of $T_c$) for the P and AP cases is suppressed as $e^{-2k_h d_F}$
(the short-range components go from the S/F1 to F1/F2 interface and back),
the changes in $W$ due to noncollinear magnetizations contain the same exponential,
$e^{-2k_h d_F}$. Of course, the influence of the long-range triplet components
is contained in prefactors but no long-ranged exponential (with $k_\omega$
instead of $k_h$) appears in $W$, because $W$ still originates from the
short-range components.

In conclusion, we have considered a mesoscopic S/F1/F2 structure composed of a superconducting
layer S, a ferromagnetic layer of arbitrary thickness F1, and a ferromagnetic half-space F2. We have demonstrated that the structure exhibits
different relations between the critical temperatures
in the parallel and antiparallel configuration: both the standard ($T_{c}^\mathrm{P}<T_{c}^\mathrm{AP}$)
and inverse ($T_{c}^\mathrm{P} >T_{c}^\mathrm{AP}$) switching can be realized depending on the system's parameters. At the same time, our main result is
that $T_c^\mathrm{Tr}$ at noncollinear magnetizations is lower than both $T_c^\mathrm{P}$ and $T_c^\mathrm{AP}$, which makes
this system a triplet spin valve. Possible experimental observation of a nonmonotonic (like curve ``3'' in Fig.~3 or curves ``1'' and ``3'' in Fig.~4) or even reentrant (like curves ``2'' and ``4'' in Fig.~4) behavior of $T_{c}(\alpha)$ could be a signature of existence of the long-range triplet superconducting correlations \cite{Bergeret} in SF hybrid structures.

We are grateful to I.\,A.\ Garifullin and A.\,S.\ Sidorenko for discussions stimulating this study,
to O.\,V.\ Nedopekin for assistance in numerical calculations, and
to M.\,V.\ Feigel'man and V.\,V.\ Ryazanov for discussion of the results.
This work was supported by the RFBR (projects 07-02-00963-a, 09-02-12176-ofi\_m, 09-02-12260-ofi\_m, and 10-02-90014-Bel\_a),
by the NanoNed (project TCS7029),
and by the Russian Federal Agency of Education (contract NK-529P).

\end{document}